\documentclass[useAMS,usenatbib]{mn2e}
\usepackage{longtable}
\usepackage{graphicx}
\usepackage{lscape}
\usepackage{rotating}
\newcommand{\mnras}{MNRAS}

\newcommand{\apj}{ApJ}
\newcommand{\apjs}{ApJ}

\newcommand{\aap}{A\&A}

\newcommand{\nat}{Nature}
\newcommand{\memsai}{Memorie della Societa Astronomica Italiana}

\newcommand{\gppr}{\stackrel{>}{\scriptstyle \sim}}
\newcommand{\gappr}{\raisebox{-0.4ex}{$\gppr$}}

\newcommand{\Mwd}{\mbox{$M_\mathrm{wd}$}}
\newcommand{\Msec}{\mbox{$M_\mathrm{sec}$}}

\newcommand{\Msun}{\mbox{$\mathrm{M}_{\odot}$}}

\newcommand{\Teff}{\mbox{$T_{\mathrm{eff}}$}}

\newcommand{\Porb}{\mbox{$P_{\mathrm{orb}}$}}

\newcommand{\Lines}[3]{\Ion{#1}{#2}\,$\lambda\lambda$\,#3}
\newcommand{\Ion}[2]{#1{\,\scriptsize #2}}

\newcommand{\kms}{\mbox{$\mathrm{km\,s^{-1}}$}}
\title[The longest period SDSS-PCEBs] {Post-common envelope binaries
  from SDSS - XVI. Long orbital period systems and the energy budget
  of CE evolution}

\author[A.    Rebassa-Mansergas  et   al.]{A.   Rebassa-Mansergas$^1$,
  M. Zorotovic$^1$, M.R. Schreiber$^1$, B.T. G\"ansicke$^2$,\newauthor
  J.  Southworth$^3$, A.   Nebot G\'omez-Mor\'an$^4$, C.  Tappert$^1$,
  D.   Koester$^5$,  S.    Pyrzas$^2$, \newauthor  C.   Papadaki$^6$,  
  L. Schmidtobreick$^7$, A. Schwope$^8$, O. Toloza$^1$\\
$^{1}$ Departamento de F\'\i sica y Astronom\'\i a, Universidad de Valpara\'\i so, 
Avenida Gran Bretana 1111, Valpara\'\i so, Chile \\
$^{2}$ Department of Physics, University of Warwick, Coventry CV4 7AL, UK \\
$^{3}$ Astrophysics Group, Keele University, Staffordshire, ST5 5BG, UK\\
$^{4}$ CNRS, Observatoire Astronomique, 11 rue de
l'Universite, F-67000 Strasbourg, France\\
$^{5}$ Institut f\"ur Theoretische Physik und Astrophysik, University of Kiel,
24098 Kiel, Germany\\
$^{6}$ Institute of Astronomy \& Astrophysics, National Observatory of Athens, 
15236 Athens, Greece\\
$^{7}$ European Southern Observatory, Alonso de Cordova 3107, Santiago, Chile \\
$^{8}$ Astrophysikalisches Institut Potsdam, An der Sternwarte 16, D-14482 Potsdam, Germany
}

\begin{document}
\date{Accepted 2011. Received 2011; in original form 2011}
\pagerange{\pageref{firstpage}--\pageref{lastpage}} \pubyear{2011}
\maketitle

\begin{abstract}
Virtually  all   close  compact   binary  stars  are   formed  through
common-envelope (CE)  evolution. It is generally  accepted that during
this crucial  evolutionary phase a  fraction of the orbital  energy is
used to expel the envelope.  However, it is unclear whether additional
sources of energy,  such as the recombination energy  of the envelope,
play an  important role.  Here we  report the discovery  of the second
and third longest orbital period post-common envelope binaries (PCEBs)
containing white dwarf (WD) primaries, i.e.  SDSSJ\,121130.94-024954.4
($\Porb=7.818\pm0.002$\,days)       and      SDSSJ\,222108.45+002927.7
($\Porb=9.588\pm0.002$\,days), reconstruct their evolutionary history,
and discuss  the implications for  the energy budget of  CE evolution.
We find  that, despite  their long orbital  periods, the  evolution of
both   systems   can  still   be   understood  without   incorporating
recombination energy,  although at  least small contributions  of this
additional  energy  seem  to   be  likely.   If  recombination  energy
significantly contributes to the  ejection of the envelope, more PCEBs
with relatively  long orbital periods  ($\Porb\gappr$1-3\,d) harboring
massive WDs ($\Mwd\gappr0.8\Msun$) should exist.

\end{abstract}


\label{firstpage}

\section{Introduction}
\label{s-intro}

Some of the  most interesting objects in our  Galaxy are close compact
binary  stars, such  as cataclysmic  variables (CVs),  low  mass X-ray
binaries, or  double degenerate white  dwarf (WD) binaries.   The vast
majority of  close compact binaries form through  common envelope (CE)
evolution occurring  when the  more massive star  of the  initial main
sequence binary fills  its Roche-lobe on the first  giant branch (FGB)
or on the asymptotic giant  branch (AGB). This may trigger dynamically
unstable mass transfer causing the giant's envelope to engulf its core
(the  future compact  object) and  the main-sequence  companion.  Drag
forces transfer  orbital energy and  angular momentum from  the binary
orbit  to   the  envelope,  reducing  the   binary  separation,  until
eventually  the  envelope  is  expelled  and a  short  orbital  period
post-common envelope binary (PCEB)  consisting of a compact object and
a main-sequence companion is exposed.

A commonly used method to predict the outcome of binary star evolution
and to theoretically investigate close compact binary star populations
are  Binary  Population  Synthesis   (BPS)  studies  which  have  been
performed     e.g.     for     Supernova    Type     Ia    progenitors
\citep{han+podsiadlowski04-1},     short      gamma     ray     bursts
\citep{belczynskietal06-1}, or  Galactic WD plus  main sequence (WDMS)
binaries \citep{willems+kolb04-1,davisetal10-1}.   However, in current
BPS  models CE evolution  is commonly  approximated by  a parametrized
energy  equation, i.e.  a  fraction of  the available  orbital energy,
known as  the CE efficiency ($\alpha_\mathrm{CE}$), is  equated to the
binding  energy  of  the envelope  \citep{paczynski76-1,  webbink84-1,
  iben+tutukov86-1,  iben+livio93-1}.  While  recent  observational as
well as theoretical results indicate rather small efficiencies for the
use    of   orbital    energy,    i.e.    $\alpha_\mathrm{CE}\sim0.25$
\citep{zorotovicetal10-1, ricker+taam12-1}, it remains unclear if, and
to what  extent, additional energy  sources play an important  role in
unbinding the envelope.

On   the   one   hand,   the   long  orbital   period   PCEB   IK\,Peg
\citep{landsmanetal93-1,  vennesetal98-1}  and  perhaps also  the  two
symbiotic    systems    T\,CrB    \citep{webbink76-1}   and    RS\,Oph
\citep{livioetal86-1} have been claimed to provide direct evidence for
additional       energy       contributions      \citep{davisetal10-1,
  zorotovicetal10-1} and  atomic recombination is  often considered to
be the  most promising candidate  \citep[e.g.][]{webbink07-1}.  On the
other hand,  \citet{soker+harpaz03-1} argue that  recombination energy
cannot significantly contribute to  the ejection process as, according
to  them, the  opacity in  the envelope  is too  small and  the energy
provided  by  recombination  should   be  radiated  away  rather  than
accelerating the gas.
 
\begin{figure}
\includegraphics[angle=-90,width=\columnwidth]{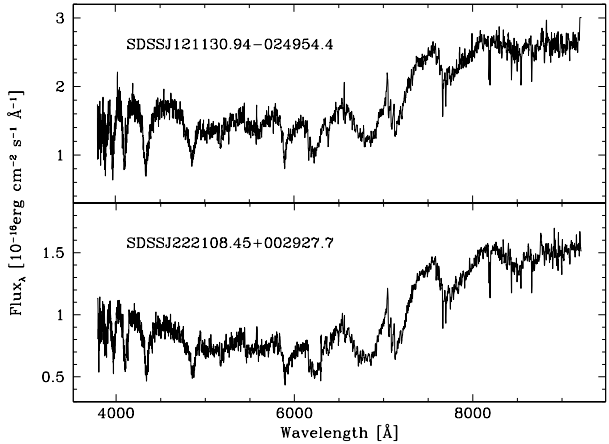} 
\caption{\label{f-spec} SDSS spectra of SDSSJ\,1211-0249 and
  SDSSJ\,2221+0029.}
\end{figure}

During  the last  few years  we have  successfully identified  a large
number of  PCEBs among WDMS  binaries discovered by the  Sloan Digital
Sky          Survey          \citep[SDSS,][]{adelman-mccarthyetal08-1,
  abazajianetal09-1}, and  measured the orbital periods  of 58 systems
\citep{schreiberetal10-1,                    rebassa-mansergasetal11-1,
  nebotetal11-1}. So far  we have found not a  single system providing
additional direct  evidence for recombination energy  to be important.
As a continuation of this large scale project, we here present orbital
period  measurements   of  the  PCEBs   SDSSJ\,121130.94-024954.4  and
SDSSJ\,222108.45+002927.7      (hereafter     SDSSJ\,1211-0249     and
SDSSJ\,2221+0029, see their  SDSS spectra in Figure\,\ref{f-spec}) and
find these two  systems to be the longest orbital  period PCEBs in our
sample, and  currently the second  and third longest WDMS  PCEBs known
after  IK\,Peg.  We  discuss the  implications of  these  findings for
theories  of CE  evolution with  particular emphasis  on  the possible
contributions  of recombination  energy  to the  energy  budget of  CE
evolution.

\section{Observations}
\label{s-obs}

We start with a brief summary of the performed spectroscopic follow-up
observations     of     SDSSJ\,1211-0249     and     SDSSJ\,2221+0029.
Instrumentation,  data   reduction  and  calibration   procedures  are
identical to  those described in \citet{nebotetal11-1}.  A  log of the
observations is provided in Table\,\ref{t-log}.

\subsection{SDSSJ\,1211-0249}

\begin{table}
\setlength{\tabcolsep}{2ex}
\caption{\label{t-log}  Log  of  the  observations. Provided  are  the
  telescopes and instruments used,  and the observing dates (observing
  periods  are   provided  for  the   GS  and  VLT   telescopes).  The
  corresponding NTT ESO program ID is 082.D-507(B).}
\begin{center}
\begin{small}
\begin{tabular}{cccc}\hline\hline
Object & Telescope & Instr.  & Date or \\
SDSSJ  &           &         & Obs. Period  \\
\hline
       1211-0249   & GS      & GMOS  & 2008\,A and \,B \\
                   & NTT     & EFOSC & Mar. 17-25 2009\\
                   & M.Baade & IMACS & May 14-16 2009\\  
                   & M.Baade & IMACS & Dec. 26-29 2009\\ 
                   & VLT     & FORS2 & 085.D-0974(A) (2010) \\
                   & VLT     & FORS2 & 087.D-0721(A) (2011) \\  
\hline   
       2221+0029   & VLT     & FORS2 & 080.D-0407(A) (2007)\\
                   & WHT     & ISIS  & 5-10 Jul. 2008 \\
                   & CA3.5   & TWIN  & 25-28 Jul. 2008 \\
                   & CA3.5   & TWIN  & 24-25 Sep. 2009 \\
                   & VLT     & FORS2 & 085.D-0974(A) (2010) \\
                   & VLT     & FORS2 & 087.D-0721(A) (2011) \\
\hline
\end{tabular}
\end{small}
\end{center}
\end{table}

SDSSJ\,1211-0249  was identified  as a  PCEB  by \citet{nebotetal11-1}
based  on   three  \Lines{Na}{I}{8183.27,8194.81}  absorption  doublet
radial velocities measurements from GMOS spectra taken at Gemini South
(GS)  during  the  semesters  2008\,A  and\,B.   Additional  follow-up
spectroscopy   aiming    to   determine   the    orbital   period   of
SDSSJ\,1211-0249 was  performed at the New  Technology Telescope (NTT)
equipped with EFOSC  during eight consecutive nights. We  took a total
of  18   spectra  providing   radial  velocities  with   rather  large
uncertainties  ($\sim20-30\,  \kms$) due  to  relatively poor  weather
conditions.  This first  data-set revealed  long-term  radial velocity
variations for SDSSJ\,1211-0249. Additional follow-up spectroscopy was
performed at Magellan/Baade armed with  IMACS during two runs of three
and  four nights  respectively,  resulting in  five additional  radial
velocities  revealing a  promising  orbital period  estimate of  about
seven days.  However several aliases resulting  from integer multiples
of  a  day did  not  allow a  definite  determination  of the  orbital
period. Finally, service mode observations at the Very Large Telescope
(VLT)  UT\,1 equipped  with FORS2  in periods  85 and  87  provided 14
additional radial velocities spanning the entire semesters which broke
the  alias degeneracy and  allowed to  accurately measure  the orbital
period.

\subsection{SDSSJ\,2221+0029}

Based  on two  spectra obtained  with  VLT/FORS2 during  period 80  we
discovered    the   close    binary    nature   of    SDSSJ\,2221+0029
\citep{nebotetal11-1}. A  first attempt to measure  the orbital period
was  performed with  ISIS mounted  at the  William  Herschel Telescope
(WHT),  where we  obtained  six  spectra. Given  the  long term  trend
revealed by the  radial velocities derived from these  WHT spectra, we
obtained  seven additional spectra  during two  observing runs  at the
3.5\,m telescope at Calar  Alto (CA\,3.5) equipped with TWIN. However,
as in the case of SDSSJ\,1211-0249, the short time span of our visitor
mode observations provided multiple  choices for the orbital period of
SDSSJ\,2221+0029.   We hence  obtained service  mode  observations (20
spectra)  at the  VLT/FORS2  during  periods 85  and  87 that  finally
allowed   to   unambiguously   determine   the   orbital   period   of
SDSSJ\,2221+0029.

\section{Orbital Periods}
\label{s-porb}

The  data described  in  Section\,\ref{s-obs} allow  us to  accurately
determine    the    orbital    periods   of    SDSSJ\,1211-0249    and
SDSSJ\,2221+0029. Radial  velocities were  measured in all  cases from
the  \Lines{Na}{I}{8183.27,8194.81} absorption  doublet,  in the  same
fashion  as  described  in  \citet[][]{rebassa-mansergasetal08-1}  and
\citet{schreiberetal08-1}. The measured radial velocities are given in
Table\,\ref{t-rvs}.

\begin{figure}
\includegraphics[angle=-90,width=\columnwidth]{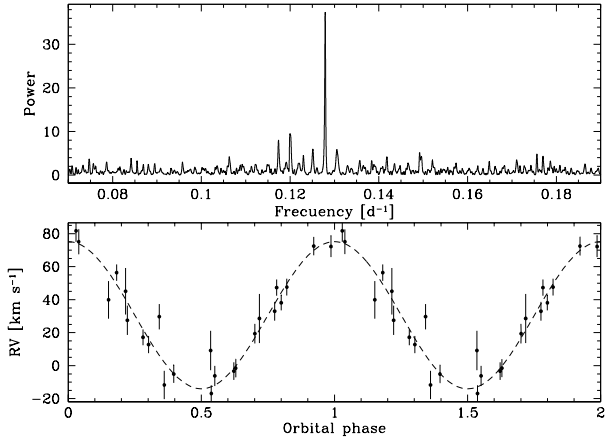}
\caption{\label{f-curve1211}  Top panel:  \textsf{ORT/TSA} periodogram
  obtained  from  the  radial  velocity data  of  SDSSJ\,1211-0249  in
  Table\,\ref{t-rvs}.   A   clear  peak  at   0.128\,d$^{-1}$  can  be
  seen. Bottom panel: the radial velocity curve folded over the period
  provided by the periodogram in the top panel.}
\end{figure}

\begin{figure}
\includegraphics[angle=-90,width=\columnwidth]{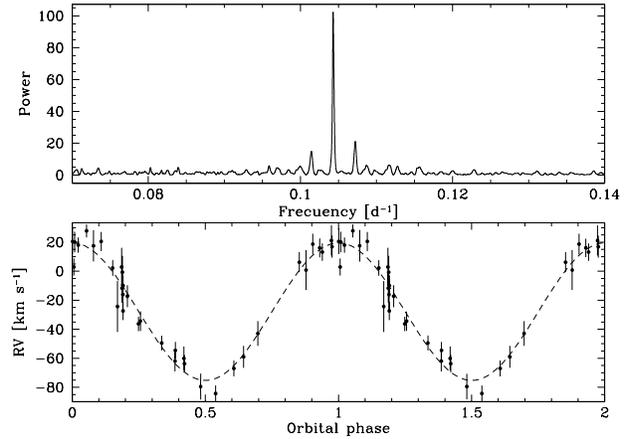}
\caption{\label{f-curve2221}  Top panel:  \textsf{ORT/TSA} periodogram
  obtained  from  the  radial  velocity data  of  SDSSJ\,2221+0029  in
  Table\,\ref{t-rvs}.   A   clear  peak  at   0.104\,d$^{-1}$  can  be
  seen. Bottom panel: the radial velocity curve folded over the period
  provided by the periodogram in the top panel.}
\end{figure}

A   \citet{scargle82-1}  periodogram   calculated   from  the   radial
velocities of  SDSSJ\,1211-0249 to investigate the  periodic nature of
the  velocity variations  contained a  number  of aliases  due to  the
sampling  pattern  of  the   visitor  mode  observations.   Using  the
\textsf{ORT/TSA} command in \textsf{MIDAS}, which folds and phase-bins
the data  using a grid of trial  periods and fits a  series of Fourier
terms      to      the      folded     radial      velocity      curve
\citep{schwarzenberg-czerny96-1}, we produced the periodogram shown in
the top panel of  Figure\,\ref{f-curve1211} which reveals a clear peak
at 0.128\,d$^{-1}$.  The same  method applied to the radial velocities
of  SDSSJ\,2221+0029  yields  a  periodogram  with  a  clear  peak  at
0.104\,d$^{-1}$ (top panel of Figure\,\ref{f-curve2221}).

To obtain a definite value  for the orbital periods we finally carried
out sine-fits of the form
\begin{equation}
\label{e-fit}
V_\mathrm{r} =
K_\mathrm{sec}\,\sin\left[\frac{2\pi(t-T_0)}{P_\mathrm{orb}}\right]
+\gamma
\end{equation}
to  the radial  velocity data  sets,  where $\gamma$  is the  systemic
velocity,  $K_\mathrm{sec}$ is the  radial velocity  semi-amplitude of
the companion star,  $T_0$ is the time of  inferior conjunction of the
secondary star, and $P_\mathrm{orb}$ is the orbital period. We adopted
the frequency corresponding to the strongest peaks in the periodograms
as  the initial orbital  period. The  parameters resulting  from these
fits are reported in Table\,~\ref{t-param} with the orbital periods of
SDSSJ\,1211-0249   and  SDSSJ\,2221+0029  being   7.818$\pm$0.002  and
9.588$\pm$0.002  days  respectively.  These  are  the longest  orbital
periods measured so far in our survey.

\begin{table*}
\setlength{\tabcolsep}{0.6ex}
\caption{\label{t-rvs}  \Ion{Na}{I} radial  velocities (RV)  and their
  errors  (RVe) measured  for  SDSSJ\,1211-0249 and  SDSSJ\,2221+0029.
  Heliocentric  Julian  dates  (HJD)  are also  provided.  The  radial
  velocities are given in $\kms$.}
\begin{center}
\begin{small}
\begin{tabular}{ccccccccccccccc}\hline\hline
HJD & RV & RVe & HJD & RV & RVe & HJD & RV & RVe & HJD & RV & RVe & HJD & RV & RVe \\
\hline
SDSSJ1211-0229 & &  & & & & & & & & & & & & \\
 2454510.792206 &   72.5 &   5.6 & 2454642.555408 &   33.0 &   5.7 & 2454644.539962 &   81.7 &   6.6 & 2454911.808350 &   45.1 &  14.1 & 2454915.737374 &   28.6 &  14.9 \\
 2454966.586407 &   27.5 &   8.9 & 2454967.523802 &   29.7 &   7.7 & 2454967.671370 &  -11.7 &   8.6 & 2455192.758500 &   39.9 &  11.4 & 2455195.755007 &    9.2 &  11.9 \\
 2455287.587077 &   17.2 &   4.8 & 2455289.592162 &  -16.9 &   5.0 & 2455291.642618 &   38.1 &   4.6 & 2455293.515465 &   75.1 &   7.5 & 2455295.568011 &   12.8 &   5.2 \\
 2455297.512659 &   -6.2 &   6.1 & 2455299.628209 &   47.6 &   5.1 & 2455306.504832 &   19.4 &   5.9 & 2455660.552464 &   89.4 &  11.5 & 2455665.519683 &   -3.7 &   6.8 \\
 2455671.572157 &   -5.5 &   5.9 & 2455674.596450 &   49.1 &   5.1 & 2455677.712572 &   57.5 &   5.2 & 2455704.665357 &    3.7 &  12.6 &                &        &       \\
\hline   
SDSSJ2221+0029 & &  & & & & & & & & & & & & \\
 2454386.661042 &   27.8 &   4.5 & 2454387.598545 &    2.2 &   5.3 & 2454653.712600 &    8.7 &   7.2 & 2454654.647388 &    0.5 &   8.7 & 2454654.719920 &    0.1 &   7.1 \\
 2454655.676768 &    0.5 &   6.6 & 2454656.622337 &   -7.2 &   7.6 & 2454658.677219 &   -3.7 &   5.9 & 2455329.885273 &   -0.1 &   8.3 & 2455334.874227 &    3.2 &   6.0 \\
 2455344.812435 &    6.9 &   7.6 & 2455346.823441 &    2.9 &  13.2 & 2455346.834686 &   -1.7 &  10.0 & 2455346.851469 &    0.7 &  11.3 & 2455346.862695 &   -6.1 &  11.3 \\
 2455354.846936 &    8.0 &   5.3 & 2455357.851973 &   -9.5 &   5.0 & 2455359.794378 &   -4.3 &   5.5 & 2455360.805659 &   -9.0 &   7.6 & 2455382.720068 &    6.1 &   6.4 \\
 2455385.783274 &   -6.3 &   5.0 & 2455399.671028 &   -2.9 &   8.5 & 2455699.895126 &    0.4 &   6.1 & 2455711.878306 &   -5.4 &   9.2 & 2455720.842993 &   -5.4 &   9.0 \\
 2455724.834694 &   -8.9 &   5.8 & 2455736.779456 &    4.4 &   6.5 & 2455741.903011 &   -7.9 &   5.9 & 2454672.643400 &    0.8 &  13.7 & 2454673.556708 &    1.2 &  10.4 \\
 2454674.577014 &    7.5 &  10.7 & 2454675.442195 &   -4.3 &  17.5 & 2454675.641213 &   -7.4 &   6.3 & 2455099.432743 &   -2.0 &   7.1 & 2455100.353820 &   -9.5 &   8.8 \\
\hline
\end{tabular}
\end{small}
\end{center}
\end{table*}

\section{Binary parameters}
\label{s-param}

\begin{figure*}
\includegraphics[angle=-90,width=\columnwidth]{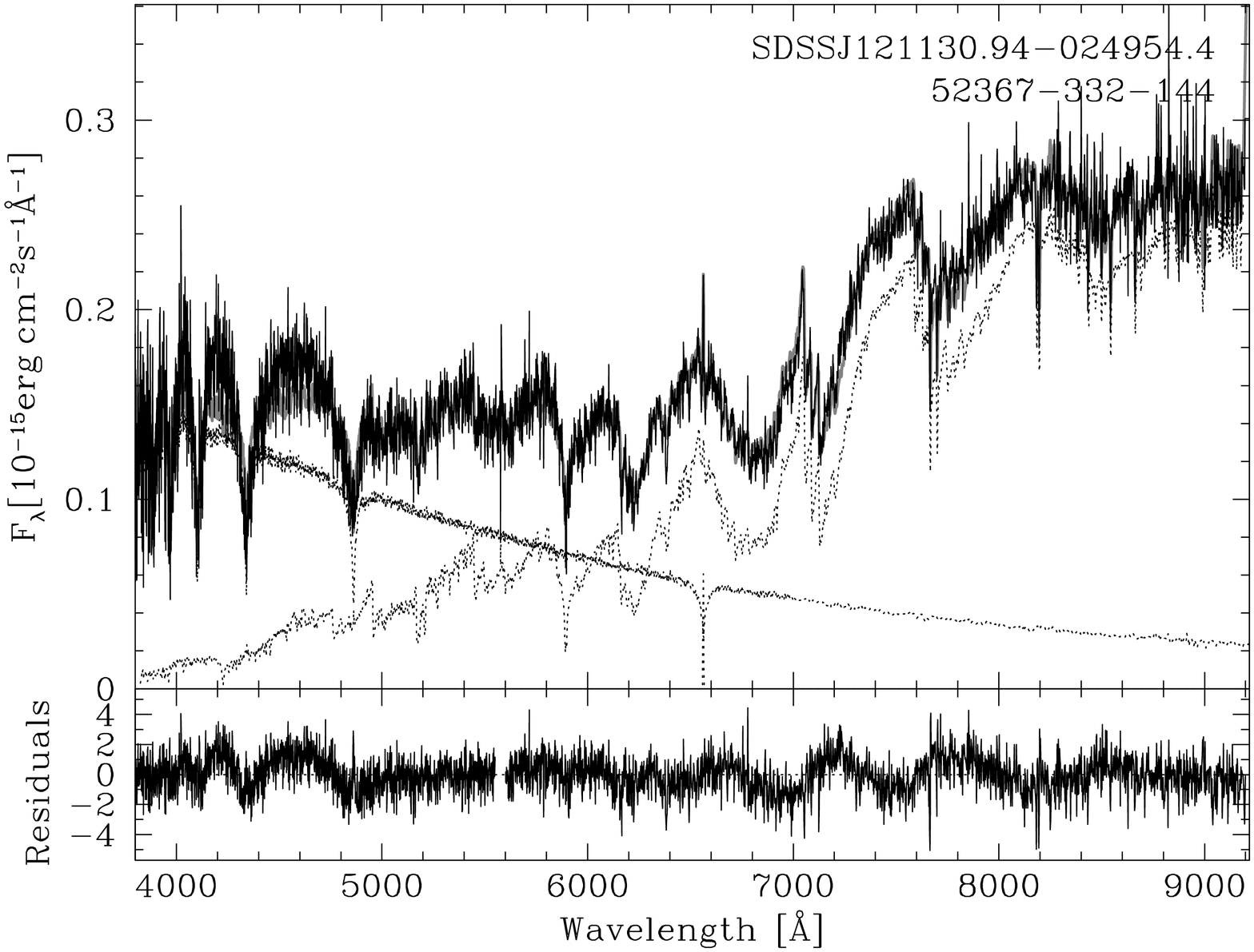}
\includegraphics[angle=-90,width=\columnwidth]{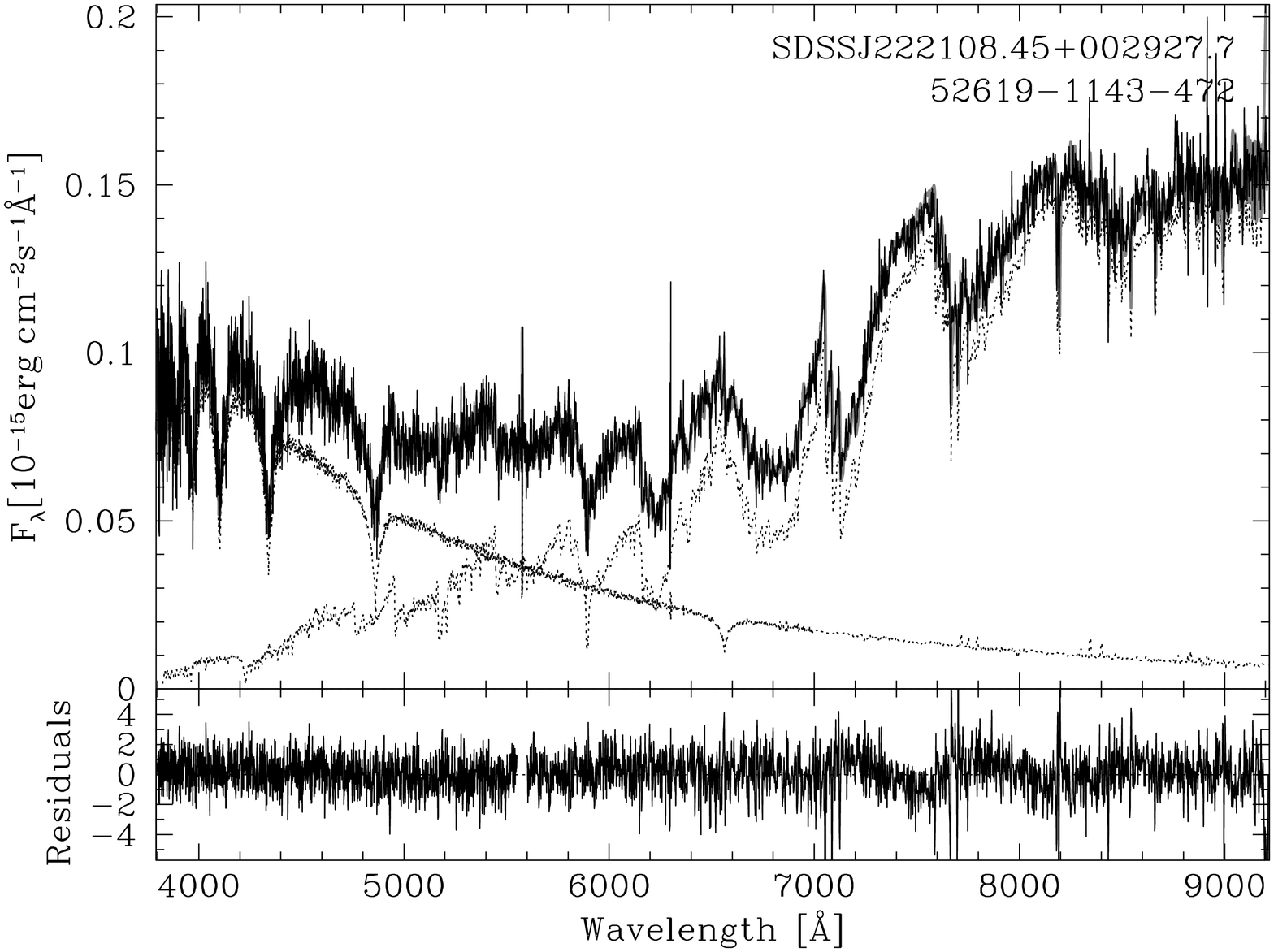}
\caption{Two-component fit  to the spectra  of SDSSJ\,1211+0249 (left)
  and SDSSJ\,2221+0029 (right). The top panels show the spectra of the
  objects as  a solid black lines  and the two  templates, white dwarf
  and M-dwarf, as  dotted lines. The bottom panel  shows the residuals
  from  the  fit.  SDSS  MJD,   PLT  and  FIB  identifiers  are  also
  indicated.}
  \label{f-rdfits} 
\end{figure*}

\begin{figure*}
\includegraphics[width=84mm]{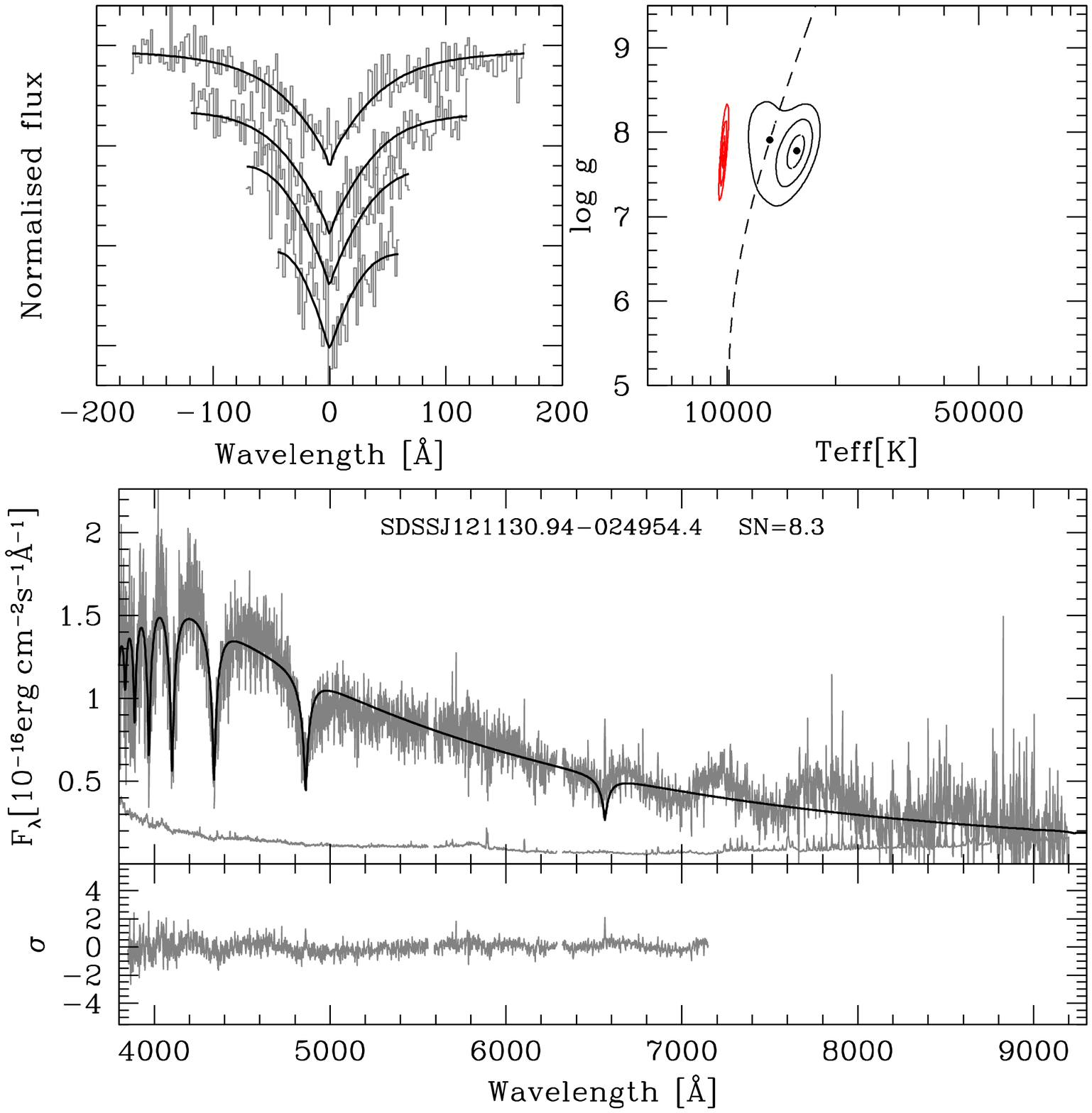}
\includegraphics[width=84mm]{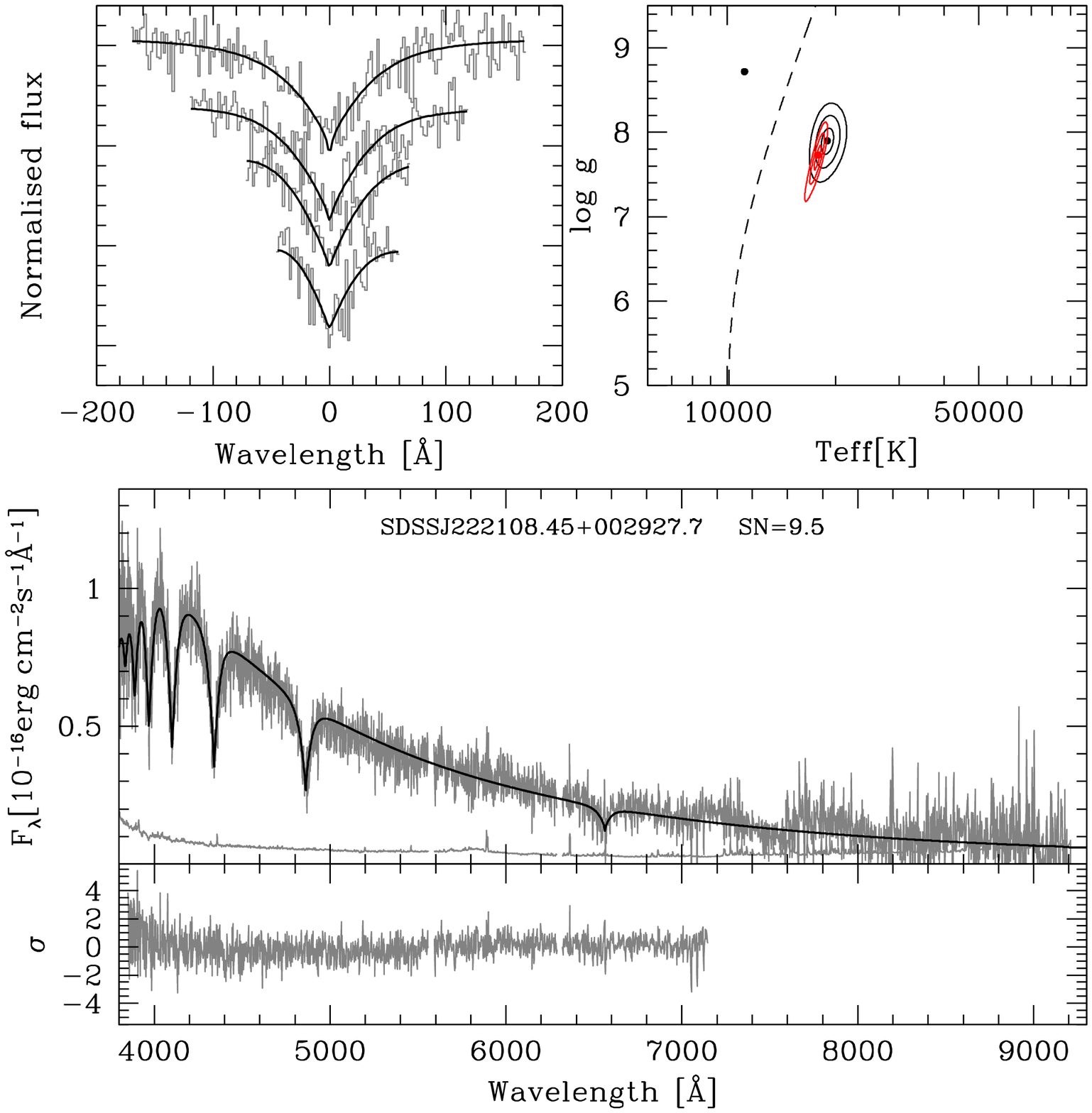}
\caption{Spectral model fit to the white dwarf in SDSSJ\,1211+0249 and
  SDSSJ\,2221+0029,  obtained after  subtracting the  best-fit M-dwarf
  templates  (see Figure\,\ref{f-rdfits}).  Top left  panels: best-fit
  (black    lines)    to    the    observed    $\mathrm{H}\beta$    to
  $\mathrm{H}\epsilon$ (gray lines, top  to bottom) line profiles. The
  model  spectra and  observations have  been normalised  in  the same
  way. Top  right panels:  1, 2 and  3\,$\sigma$ contour plots  in the
  \Teff-$\log  g$ plane.  The black  contours refer  to the  best line
  profile fit, the red ones (which collapse into a dot on the scale of
  the  plot) to  the fit  of the  spectral range  3850--7150\,\AA. The
  dashed  line indicates the  occurrence of  maximum $\mathrm{H}\beta$
  equivalent  width.  The  best  ``hot''  and  ``cold''  line  profile
  solutions are  indicated by  black dots, while  the best fit  to the
  whole spectrum by a red one. Bottom panels: the residual white dwarf
  spectra  resulting from  the spectral  decomposition and  their flux
  errors (gray lines) along with the best-fit white dwarf model (black
  lines)  in  the  3850--7150\,\AA\  wavelength range  (top)  and  the
  residuals of the fit (gray line, bottom).}
  \label{f-wdfits} 
\end{figure*}

We provide in this Section the binary (orbital and stellar) parameters
of the two PCEBs studied  in this work.  The WD effective temperatures
($\Teff_\mathrm{(WD)}$),  surface  gravities ($\log  g_\mathrm{(WD)}$)
and masses  ($M_\mathrm{wd}$), as well as the  secondary star spectral
types   (Sp$_\mathrm{sec}$),   masses   and  radii   ($M_\mathrm{sec},
R_\mathrm{sec}$)  are  obtained  following  the  decomposition/fitting
technique  described in  \citet{rebassa-mansergasetal07-1}.   In brief
this routine  follows a two-step procedure.  First,  the SDSS spectrum
is fitted  with a  two-component model, and  the spectral type  of the
secondary  star is  determined (Figure\,\ref{f-rdfits}).   Second, the
best-fit M-dwarf is subtracted and  the residual WD spectrum is fitted
with a  model grid of DA  WDs \citep{koester10-1} to  determine the WD
effective  temperature and  surface  gravity (Figure\,\ref{f-wdfits}).
From  an  empirical spectral  type-radius-mass  relation for  M-dwarfs
\citep{rebassa-mansergasetal07-1} and  a mass-radius relation  for WDs
\citep{bergeronetal95-2,fontaineetal01-1}  we then calculate  the mass
and radius of the secondary star and the WD respectively.

For the majority of SDSS PCEBs the spectroscopic decomposition results
in an  uncertainty of the  spectral type of $\pm0.5$  spectral classes
\citep{rebassa-mansergasetal10-1, rebassa-mansergasetal12-1}, and this
applies as  well to SDSS\,J2221+0029.   However, the SDSS  spectrum of
SDSS\,J1211-0249  suffers  from   low-frequency  structure  (there  is
substantial  structure left  in  the residual  WD  spectrum after  the
decomposition) that  results in  a substantially larger  uncertainty in
the  determination  of  the   spectral  type  of  the  companion,  and
correspondingly  larger uncertainties  in the  white  dwarf parameters
(see      left      panels      of     Figure\,\ref{f-wdfits}      and
Table\,\ref{t-param}).    For    SDSS\,J2221+0029,    $M_\mathrm{wd}$,
$\Teff_\mathrm{(WD)}$,  $\log  g_\mathrm{(WD)}$, $M_\mathrm{sec}$  and
Sp$_\mathrm{sec}$  are obtained by  averaging the  fit results  of two
independent SDSS spectra, and  the uncertainties are the corresponding
standard deviations.  For  SDSS\,J1211-0249, we average the parameters
over  two  possible solutions  for  the  spectral decomposition  using
either an  M2 or  M3 template, and  determine the  uncertainties again
from the corresponding standard deviations.

To calculate the binary inclinations we use Kepler's third law,
\begin{equation}
\frac{(M_\mathrm{wd}\sin i)^3}{(M_\mathrm{wd}+M_\mathrm{sec})^2}
=\frac{\Porb K_\mathrm{sec}^3}{2\pi G}
\end{equation}
\noindent rewritten as
\begin{equation}
\sin i = \frac{K_\mathrm{sec}}{M_\mathrm{wd}} \left(\frac{\Porb}
{2\pi G}\right)^{1/3}(M_\mathrm{wd}+M_\mathrm{sec})^{2/3},
\label{e-incl}
\end{equation}
\noindent
with  the  orbital  periods   and  semi-amplitude  velocities  of  the
companions  $K_\mathrm{sec}$ as  determined  in Section\,\ref{s-porb},
and  the stellar  masses as  obtained from  the analysis  of  the SDSS
spectra     outlined    above.      The     well    known     relation
$M_\mathrm{sec}/M_\mathrm{wd}     =    K_\mathrm{wd}/K_\mathrm{sec}=q$
provides an estimate of the expected semi-amplitude velocity of the WD
$K_\mathrm{wd}$.   Finally, estimates of  the orbital  separations and
Roche lobe radii of the secondary stars $R_\mathrm{L\mathrm{sec}}$ are
obtained     from     Kepler's     third    law     and     Eggleton's
(\citeyear{eggleton83-1}) expression
\begin{equation}
R_\mathrm{L\mathrm{sec}} = \frac{a\,0.49\,q^{2/3}}
{0.6\,q^{2/3} + \ln(1+q^{1/3})}
\label{e-rl}
\end{equation}
respectively.    The   complete   sets   of  binary   parameters   for
SDSSJ\,1211-0249     and     SDSSJ\,2221+0029     are     given     in
Table\,\ref{t-param}.

\begin{table}
\caption{\label{t-param}     Binary     parameters    obtained     for
  SDSSJ\,1211-0249     and     SDSSJ\,2221+0029.      $M_\mathrm{wd}$,
  $M_\mathrm{sec}$, $R_\mathrm{sec}$,  spectral type of  the companion
  Sp$_\mathrm{sec}$, $\Teff_\mathrm{(WD)}$\, and $\log g$ are obtained
  following    the   decomposition/fitting   routine    described   in
  \citet{rebassa-mansergasetal07-1}.   The orbital  period  \Porb, the
  secondary  star semi-amplitude  $K_\mathrm{sec}$,  and the  systemic
  velocity       $\gamma_\mathrm{sec}$      are       measured      in
  Section\,\ref{s-porb}.   Estimates of  the  orbital separation  $a$,
  mass ratio $q$, white dwarf semi-amplitude velocity $K_\mathrm{wd}$,
  secondary   Roche  lobe   radius   R$_\mathrm{L_\mathrm{sec}}$,  and
  inclination   are    obtained   from   the    equations   given   in
  Section\,\ref{s-param}.}
\begin{flushleft}
\begin{center}
\setlength{\tabcolsep}{0.9ex}
\begin{tabular}{llcccccc}
\hline\hline
\noalign{\smallskip}
 & &\multicolumn{3}{c}{SDSS\,J1211--0229}  & \multicolumn{3}{c}{SDSS\,J2221+0029}\\
\hline
$M_\mathrm{wd} [M_\odot]$                  &&    0.52&$\pm$&0.07    &   0.54&$\pm$&0.03\\
$M_\mathrm{sec} [M_\odot]$                 & &   0.41&$\pm$&0.05  &   0.38&$\pm$&0.07\\
$q$                                      &&    0.79&$\pm$&0.15    &   0.70&$\pm$&0.15 \\
$a [R_\odot]$                             &&    16.2&$\pm$&0.5    &   18.5&$\pm$&0.5 \\
\Porb\ [d]                               &&  7.818&$\pm$&0.002  & 9.588&$\pm$&0.002\\
$\gamma_\mathrm{sec} [\kms]$              &&     30&$\pm$&2       &  -29&$\pm$&2    \\
$K_\mathrm{sec} [\kms]$                   &&     44&$\pm$& 3     &   49&$\pm$& 2  \\
$K_\mathrm{wd} [\kms]$                    &&     35&$\pm$& 7     &    34&$\pm$& 7 \\
Sp$_\mathrm{sec}$                          &&     M2.5&$\pm$& 1   &   M3&$\pm$& 0.5\\
$R_\mathrm{sec} [R_\odot]$                 &&   0.42&$\pm$&0.05   &  0.39&$\pm$&0.08\\
%$R_\mathrm{wd} [R_\odot]$                 &&   0.014&$\pm$&0.004  &  0.014&$\pm$&0.001\\
$R_\mathrm{sec}/R_\mathrm{L\mathrm{sec}}$  & &   0.07&$\pm$&0.01    &   0.06&$\pm$&0.01    \\
$i [^\circ]$                               &&     49&$\pm$& 7     &    58&$\pm$& 7    \\
$\Teff_\mathrm{(WD)}$ [K]                   &&  13130&$\pm$&860   & 18440&$\pm$&150 \\
$\log g_\mathrm{(WD)}$                      &&    7.84&$\pm$&0.13    &   7.85&$\pm$&0.06  \\
\hline
\noalign{\smallskip}
\end{tabular}
\end{center}
\end{flushleft}  
\end{table}

\section{Discussion}
\label{s-disc}

We have presented in the previous Sections the discovery of the second
and  third longest  orbital period  (detached) PCEBs  containing  a WD
primary. In  what follows we  reconstruct the evolutionary  history of
both  systems and  discuss implications  for our  understanding  of CE
evolution.

\subsection{The evolution of SDSSJ\,1211-0249 and SDSSJ\,2221+0029}
\label{s-evol}

Having at hand the orbital periods, the stellar mass estimates of both
components and the WD  effective temperatures allows us to reconstruct
the evolutionary history  of SDSSJ\,1211-0249 and SDSSJ\,2221+0029 and
predict   their   future   following   \citet{zorotovicetal11-1}   and
\citet{schreiber+gaensicke03-1}  respectively.  First,  we interpolate
the       cooling      tracks       of       \citet{wood95-1}      and
\citet{althaus+benvenuto97-1}  to determine  the cooling  age  of both
systems. Second,  we derive the  orbital period at  the end of  the CE
phase  using the  the most  up-to-date version  of  disrupted magnetic
braking    \citep[][including    the    normalization   provided    by
  \citealt{davisetal08-1}]{hurleyetal02-1}.    Third,   we   use   the
single-star   evolution  (SSE)   code  of   \citet{hurleyetal00-1}  to
reconstruct the  CE phase for a  given value of the  CE efficiency and
obtain the orbital and stellar parameters prior to CE evolution.

We  here  follow   \citet{zorotovicetal10-1}  and  \emph{assume}  that
recombination energy  contributes to  expelling the envelope  with the
same    efficiency     as    the    orbital     energy    (given    by
$\alpha_{\mathrm{CE}}$) and take into account the uncertainties in the
stellar  component masses  and WD  effective temperatures.   We obtain
solutions  for rather  large  ranges  of the  CE  efficiency for  both
systems that are  given together with the resulting  range of possible
parameters and  evolutionary time scales  of the progenitor  system in
Table\,\ref{tab:reco}.  As  outlined in the  introduction, there seems
to  be some  evidence  for a  relatively  small CE  efficiency and  we
therefore  additionally  provide  the progenitor  parameters  assuming
$\alpha_{\mathrm{CE}}=0.25$      in     Table\,\ref{tab:reco}     (the
uncertainties of  the stellar masses and WD  effective temperature are
not  considered  here).  This  complements  the  results presented  in
\citealt{zorotovicetal11-1}  (their  Table\,3).  Given  that  magnetic
braking is not  efficient in long orbital period  systems, the current
orbital periods  (Table\,\ref{t-param}) are nearly  identical to those
at  the  end  of  the  CE  phase.  The  masses  of  the  reconstructed
progenitors  for both  objects are  similar, which  is  not surprising
since the available estimates of  the stellar masses are quite similar
too.

Since   the   current   orbital   periods  of   SDSSJ\,1211-0249   and
SDSSJ\,2221+0029 are very long  for PCEBs, the stellar components will
not  be close  enough to  trigger the  second phase  of  mass transfer
before  the secondaries  in both  systems  evolve away  from the  main
sequence.   Both secondaries  will  therefore fill  their Roche  lobes
during the FGB transforming SDSSJ\,1211-0249 and SDSSJ\,2221+0029 into
symbiotic  systems with stable  mass transfer  from a  red giant  to a
white dwarf. This is supposed to  happen in many Hubble times when the
orbital periods have shrunk to $\sim1-1.5$\,days.

\begin{table}
\caption{\label{tab:reco}   Applying   the  reconstruction   algorithm
  described  in  \citet{zorotovicetal11-1}  we determine  the  orbital
  periods at  the end of  CE evolution $P_{\mathrm{CE}}$,  the initial
  mass of the  primary $M_{\mathrm{1,o}}$, the mass of  the primary at
  the  onset of  CE evolution  $M_{\mathrm{1,CE}}$,  the corresponding
  orbital separation  $a_{\mathrm{i}}$, the main  sequence lifetime of
  the  primary  $t_{\mathrm{evolv}}$,  the   cooling  age  of  the  WD
  $t_{\mathrm{cool}}$,  and the time  until the  second phase  of mass
  transfer  will occur  from now  on $t_{\mathrm{sd}}$  and  since the
  progenitor  main  sequence  binary  has  formed  $t_{\mathrm{tot}}$.
  Columns   two   and   four   correspond   to  a   fixed   value   of
  $\alpha_{\mathrm{CE}}=0.25$.  Columns three and five give the entire
  range  of possible  solutions.  Note  that since  disrupted magnetic
  braking  is  very  inefficient  for  long  orbital  period  systems,
  $P_{\mathrm{CE}}$    is    nearly    identical   to    \Porb\,    in
  Table\,\ref{t-param}. }
\begin{flushleft}
\begin{center}
\setlength{\tabcolsep}{1.6ex}
\begin{tabular}{llcccccccc}
\hline\hline
\noalign{\smallskip}
                    & \multicolumn{4}{c}{SDSS\,J1211--0229} &\multicolumn{4}{c}{SDSS\,J2221+0029}\\
\hline
$\alpha_{\mathrm{CE}}$ & \multicolumn{3}{c}{0.25} & 0.03-1 & \multicolumn{3}{c}{0.25} & 0.06-1\\
\hline
$P_{\mathrm{CE}}$ [d]        & \multicolumn{3}{c}{7.820} & 7.819-7.822 & \multicolumn{3}{c}{9.589} & 9.588-9.589\\
$M_{\mathrm{1,o}} [M_\odot]$  & \multicolumn{3}{c}{1.31} & 0.98-2.35    & \multicolumn{3}{c}{1.44} & 1.31-2.35\\
$M_{\mathrm{1,CE}} [M_\odot]$ & \multicolumn{3}{c}{1.1} & 0.8-2.3     & \multicolumn{3}{c}{1.2} & 1.0-2.3\\
$a_{\mathrm{i}} [R_\odot]$    & \multicolumn{3}{c}{480.8} & 246.2-613.9 & \multicolumn{3}{c}{512.2} & 295.5-611.3 \\
$P_{\mathrm{sd}}$ [h]       & \multicolumn{3}{c}{26.0} & 24.0-28.2    & \multicolumn{3}{c}{32.0} & 28.9-36.0 \\
$t_{\mathrm{evolv}}$ [Gyr]    & \multicolumn{3}{c}{4.82} & 0.95-13.29   & \multicolumn{3}{c}{3.53} & 0.95-4.82  \\
$t_{\mathrm{cool}}$ [Gyr]    & \multicolumn{3}{c}{0.24} & 0.24-0.43    & \multicolumn{3}{c}{0.08} & 0.07-0.09 \\
$t_{\mathrm{sd}}$ [Gyr]      & \multicolumn{3}{c}{225.3} & 140.1-339.7 & \multicolumn{3}{c}{268.5} & 150.13-454.36 \\
$t_{\mathrm{tot}}$ [Gyr]     & \multicolumn{3}{c}{230.4} & 141.3-353.4 &\multicolumn{3}{c}{272.1} & 151.15-459.27\\
\hline
\noalign{\smallskip}
\end{tabular}
\end{center}
\end{flushleft} 
\end{table}

\subsection{The energy budget of CE evolution}
\label{s-budget} 

In  their  review,   \citet{iben+livio93-1}  describe  several  energy
sources apart  from orbital energy that might  contribute to expelling
the  envelope,  ranging  from  recombination  energy  to  dust  driven
winds. Since  the writing  of this review,  the energy equation  of CE
evolution  in  general, and  especially  the  potential importance  of
recombination    energy,    has    been    a    matter    of    debate
\citep[e.g.][]{dewi+tauris00-1,         webbink07-1,        xu+li10-1,
  loveridgeetal11-1, soker+harpaz03-1, zorotovicetal10-1}.

In the  previous Section  we reconstructed the  evolution of  the long
orbital    period   PCEBs   SDSSJ\,1211-0249    and   SDSSJ\,2221+0029
{\em{assuming}} that recombination energy contributes to expelling the
envelope  with the  same efficiency  as the  orbital energy  and found
large ranges of possible  solutions.  Here we investigate whether this
assumed additional energy is  a necessary ingredient to understand the
evolutionary  history of  SDSSJ\,1211-0249  and SDSSJ\,2221+0029.   To
that  end we  now reconstruct  the CE  phase of  both  systems without
considering recombination energy.  Taking into account the uncertainty
of the  measured stellar parameters, we find  possible progenitors for
both   systems    without   violating   energy    conservation,   i.e.
$\alpha_{\mathrm{CE}}=0.21-1$       for      SDSSJ\,1211-0249      and
$\alpha_{\mathrm{CE}}=0.42-1$  for   SDSSJ\,2221+0029.   We  therefore
conclude that  the existence  of the two  systems does not  confirm or
disprove whether recombination (or  any other additional) energy plays
an  important  role  during   the  CE  phase.   However,  the  current
configuration  of   SDSSJ\,2221+0029  can  only  be   explained  if  a
relatively large  fraction of the released  orbital energy contributes
to  envelope ejection,  i.e.   $\alpha_\mathrm{CE}>0.42$.  This  value
exceeds the  estimates given  in recent studies  of CE  evolution that
seem     to     converge     towards     a    CE     efficiency     of
$\alpha_{\mathrm{CE}}\sim0.25$                \citep{zorotovicetal10-1,
  ricker+taam12-1,  passyetal12-1}.   If this  is  generally true,  at
least a small  fraction of recombination energy (or  any other form of
additional energy) seems to  have contributed to the envelope ejection
in SDSSJ\,2221+0029.   Although this interpretation  appears tempting,
the fact  remains that not a  single PCEB within  the homogeneous SDSS
sample  \citep{nebotetal11-1}   provides  {\em{direct}}  evidence  for
additional sources of energy playing a role during CE evolution.

\subsection{Future perspectives}
\label{s-future}

IK\,Peg has  been highlighted  as a  key object as  it is  the longest
orbital  period system and  contains the  most massive  secondary star
among  the known  PCEBs  containing  a white  dwarf  primary.  IK  Peg
requires  extra energy  that helps  to  expel the  envelope during  CE
evolution  \citep[e.g.][]{davisetal10-1}.  Indeed,  IK\,Peg  cannot be
reconstructed unless at least a small fraction of recombination energy
is  taken  into  account  \citep{zorotovicetal10-1}.  In  contrast  to
IK\,Peg   the   two  PCEBs   discussed   here,  SDSSJ\,1211-0249   and
SDSSJ\,2221+0029,    contain   relatively   low-mass    C/O-core   WDs
(Table\,\ref{t-param}), therefore their progenitors filled their Roche
lobes early on the AGB, i.e.  when the envelope was not very extended.
Recombination energy,  however, is expected to be  most important when
the WD  progenitor radius is large  and the envelope  is loosely bound
\citep{webbink07-1}. The peculiarity of  IK\,Peg is therefore not only
its  long orbital  period but  also  the high  mass of  its WD,  which
implies that  the system entered the  CE phase when the  radius of the
primary was very large on the  AGB, a peculiarity our two PCEBs do not
share.

\begin{figure}
\includegraphics[angle=-90,width=\columnwidth]{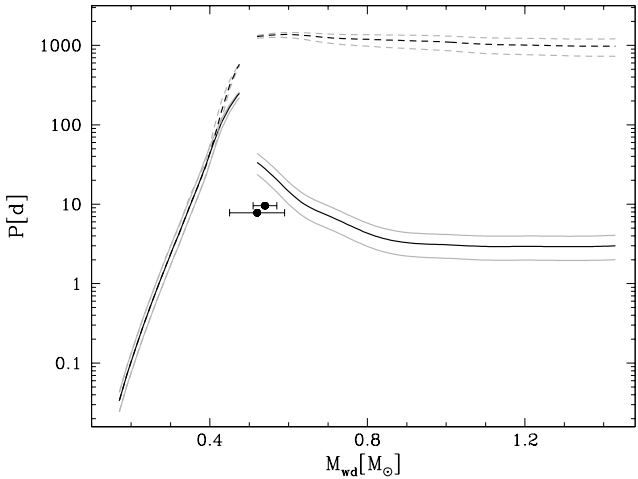} \caption{\label{fig:M1P}
  Maximum orbital period versus WD mass assuming a secondary star mass
  of $\Msec=0.4$ (black lines) $\pm0.1\Msun$ (gray lines).  The dashed
  lines correspond to the  maximum orbital period if all recombination
  energy goes  into CE  ejection, while the  solid lines  provides the
  same limit  but without  taking into account  possible contributions
  from recombination.  Any system  located between the two lines would
  provide  direct  evidence  for  the contributions  of  recombination
  energy.  The difference between both  lines is largest for high mass
  WDs as the relative  importance of recombination energy increases on
  the AGB.  So  far all PCEBs in the homogeneous  SDSS sample lie well
  below  the solid lines.   Note that  the exact  location of  the two
  lines depends  on the secondary  star mass. PCEBs with  more massive
  secondaries have  more orbital energy  available and the  limits are
  hence slightly shifted towards longer orbital periods. }
\end{figure}

To  predict which  kind  of  PCEBs would  provide  the desired  direct
evidence for  contributions of recombination  energy we once  more use
the  reconstruction algorithm described  in \citet{zorotovicetal11-1}.
As usual we assume  that the WD mass is equal to  the core mass of the
giant progenitor at the onset  of mass transfer and that the secondary
star mass remains constant during CE evolution.  For a given core mass
we  use the  SSE  code from  \citet{hurleyetal00-1}  to calculate  all
possible  progenitor masses  and their  radii.  As  the radius  of the
progenitor must have been equal to the Roche radius at the onset of CE
evolution we obtain  the initial separation for given  white dwarf and
main sequence  companion masses, leaving the final  orbital period and
the CE efficiency  as the remaining free parameters  connected via the
energy  equation.    For  each  progenitor  mass   the  solution  with
$\alpha_\mathrm{CE}  \simeq 1.0$ corresponds  to the  longest possible
final orbital  period not violating energy  conservation.  Among these
possible solutions  we finally can  select the maximum  orbital period
for a given combination of white dwarf and secondary star masses.

In Figure\,\ref{fig:M1P}  we show  the resulting PCEB  maximum orbital
period as a  function of WD mass assuming a  fixed secondary star mass
of  $\Msec=0.4\pm0.1\Msun$.   The  positions of  SDSSJ\,1211-0249  and
SDSSJ\,2221+0029 are indicated by  black solid dots.  The dashed lines
have been  obtained by assuming  that all the  available recombination
energy goes into envelope ejection while the solid lines represent the
maximum  orbital period  if the  envelope is  expelled by  the  use of
orbital  energy only.   The upper  and lower  (solid and  dashed) gray
lines    correspond   to    $\Msec=0.5\Msun$    and   $\Msec=0.3\Msun$
respectively.  The  orbital period limits increase  with the secondary
star  mass  because PCEBs  with  more  massive  secondaries have  more
orbital energy available.

Any PCEB located above the  solid line in Figure\,\ref{fig:M1P} (for a
given  secondary   star  mass)  would  provide   direct  evidence  for
contributions of additional energy sources.  Apparently, recombination
energy as  the most likely extra-energy  can only be  important on the
tip   of   the   FGB  and   on   the   AGB   (see  dashed   lines   in
Figure\,\ref{fig:M1P}).  For high  mass WDs ($\Mwd\gappr0.8\Msun$) the
range of orbital periods that would provide evidence for recombination
energy  is significantly shifted  towards shorter  (easily measurable)
orbital periods  of a few  days.  However, so  far not a  single known
PCEB  apart  from  IK\,Peg   has  a  relatively  long  orbital  period
\emph{and}  contains  a  high-mass  WD.   The seven  SDSS  PCEBs  with
accurately   determined  orbital   periods   and  stellar   parameters
containing massive  WDs ($\geq0.8\Msun$) have  orbital periods shorter
than  one  day (see  Table\,3  in \citealt{zorotovicetal11-1}).   This
might further indicate that the fraction of recombination energy going
into  envelope  ejection  is  small.  However,  further  observational
constraints  are  required  to  confirm  this  supposition.   We  have
therefore  just  started  an  observing campaign  to  measure  orbital
periods  of  additional SDSS  PCEBs  with  $\Mwd  \gappr 0.8\Msun$  to
further  constrain the  importance of  recombination energy  during CE
evolution.   The  secondary star  masses  of  the  PCEBs in  our  SDSS
follow-up project  are mostly in the  range of $\Msec\sim0.2-0.4\Msun$
which corresponds  to orbital period limits  given by the  full use of
orbital  energy  of   $\sim\,1-3$  days  (see  Figure\,\ref{fig:M1P}).
Direct evidence for additional energy, most likely from recombination,
would  be provided if  at least  one system  will be  found to  have a
period exceeding  this limit.  If in  contrast no such  system will be
detected, the contribution of recombination energy during CE evolution
is likely of minor importance.

\section{Conclusion}
\label{s-concl}

We  have   measured  the  orbital  periods   of  SDSSJ\,1211-0249  and
SDSSJ\,2221+0029  to  be   7.818$\pm$0.002  and  9.588$\pm$0.002  days
respectively.   This  makes  them  the longest  orbital  period  PCEBs
containing a  WD primary  and main sequence  companion after  the well
known record holder IK\,Peg. We reconstructed the CE evolution of both
systems  taking  into  account  and  ignoring  additional  sources  of
energy. Although no direct evidence for contributions of recombination
energy during CE  evolution is provided, it appears  plausible that at
least a small  fraction of this energy helped  expelling the envelope.
Measuring  the  orbital periods  of  more  PCEBs containing  high-mass
($\Mwd \gappr  0.8\Msun$) WDs will provide further  constraints on the
importance of recombination energy during CE evolution.
 
\section*{Acknowledgments}

ARM acknowledges financial support from  Fondecyt in the form of grant
number  3110049.  MZ acknowledges  support from  Gemini/Conicyt, grant
number  32100026.   MRS  acknowledges  support from  Milenium  Science
Initiative,  Chilean  Ministry  of  Economy,  Nucleus  P10-022-F,  and
Fondecyt (1061199).  ANGM acknowledges  support by the Centre National
d'Etudes Spatial  (CNES, ref.  60015).  This project  was supported in
part  by the  DFHG under  contract  Schw536/33-1.  We  also thank  the
anonymous referee  for his/her  suggestions that helped  improving the
quality of the paper.

The presented research is based on observations collected at the
following telescopes: the European Organisation for Astronomical
Research in the Southern Hemisphere, Chile (080.D-0407(A),
082.D-0507(B)) 085.D-0974(A), 087.D-0721(A));
the  Gemini Observatory, which is operated  by the Association of 
Universities for
Research in  Astronomy, Inc., under  a cooperative agreement  with the
NSF  on  behalf  of  the  Gemini  partnership:  the  National  Science
Foundation  (United  States), the  Science  and Technology  Facilities
Council  (United  Kingdom), the  National  Research Council  (Canada),
CONICYT   (Chile),  the   Australian  Research   Council  (Australia),
Minist\'erio  da Ci\'encia  e  Tecnologia (Brazil)  and Ministerio  de
Ciencia,   Tecnolog\'ia    e   Innovaci\'on   Productiva   (Argentina)
(GS-2008A-Q-31, GS-2008B-Q-40);  
the Magellan Baade  Telescope located
at Las  Campanas Observatory,  Chile; 
the William  Herschel Telescope,
operated on  the island of La Palma  by the Isaac Newton  Group in the
Spanish Observatorio  del Roque de  los Muchachos of the  Instituto de
Astrof\'isica  de Canarias;  
and at  the Centro  Astron\'omico Hispano
Alem\'an  (CAHA) at  Calar Alto,  operated jointly  by  the Max-Planck
Institut  f\"ur  Astronomie  and  the Instituto  de  Astrof\'isica  de
Andaluc\'ia.

%\bibliographystyle{mn_new}
%\bibliography{aamnem99,aabib}

\begin{thebibliography}{42}
\expandafter\ifx\csname natexlab\endcsname\relax\def\natexlab#1{#1}\fi

\bibitem[{{Abazajian} et~al.(2009)}]{abazajianetal09-1}
{Abazajian}, K.~N., et~al., 2009, \apjs, 182, 543

\bibitem[{{Adelman-McCarthy} et~al.(2008)}]{adelman-mccarthyetal08-1}
{Adelman-McCarthy}, J.~K., et~al., 2008, ApJS, 175, 297

\bibitem[{{Althaus} \& {Benvenuto}(1997)}]{althaus+benvenuto97-1}
{Althaus}, L.~G., {Benvenuto}, O.~G., 1997, ApJ, 477, 313

\bibitem[{{Belczynski} et~al.(2006){Belczynski}, {Perna}, {Bulik}, {Kalogera},
  {Ivanova}, \& {Lamb}}]{belczynskietal06-1}
{Belczynski}, K., {Perna}, R., {Bulik}, T., {Kalogera}, V., {Ivanova}, N.,
  {Lamb}, D.~Q., 2006, \apj, 648, 1110

\bibitem[{{Bergeron} et~al.(1995){Bergeron}, {Wesemael}, \&
  {Beauchamp}}]{bergeronetal95-2}
{Bergeron}, P., {Wesemael}, F., {Beauchamp}, A., 1995, PASP, 107, 1047

\bibitem[{{Davis} et~al.(2008){Davis}, {Kolb}, {Willems}, \&
  {G{\"a}nsicke}}]{davisetal08-1}
{Davis}, P.~J., {Kolb}, U., {Willems}, B., {G{\"a}nsicke}, B.~T., 2008, \mnras,
  389, 1563

\bibitem[{{Davis} et~al.(2010){Davis}, {Kolb}, \& {Willems}}]{davisetal10-1}
{Davis}, P.~J., {Kolb}, U., {Willems}, B., 2010, \mnras, 403, 179

\bibitem[{{Dewi} \& {Tauris}(2000)}]{dewi+tauris00-1}
{Dewi}, J.~D.~M., {Tauris}, T.~M., 2000, A\&A, 360, 1043

\bibitem[{{Eggleton}(1983)}]{eggleton83-1}
{Eggleton}, P.~P., 1983, ApJ, 268, 368

\bibitem[{{Fontaine} et~al.(2001){Fontaine}, {Brassard}, \&
  {Bergeron}}]{fontaineetal01-1}
{Fontaine}, G., {Brassard}, P., {Bergeron}, P., 2001, PASP, 113, 409

\bibitem[{{Han}(2004)}]{han+podsiadlowski04-1}
{Han}, Z.~and{Podsiadlowski}, P., 2004, MNRAS, 350, 1301

\bibitem[{{Hurley} et~al.(2000){Hurley}, {Pols}, \& {Tout}}]{hurleyetal00-1}
{Hurley}, J.~R., {Pols}, O.~R., {Tout}, C.~A., 2000, MNRAS, 315, 543

\bibitem[{{Hurley} et~al.(2002){Hurley}, {Tout}, \& {Pols}}]{hurleyetal02-1}
{Hurley}, J.~R., {Tout}, C.~A., {Pols}, O.~R., 2002, \mnras, 329, 897

\bibitem[{{Iben} \& {Tutukov}(1986)}]{iben+tutukov86-1}
{Iben}, Jr., I., {Tutukov}, A.~V., 1986, \apj, 311, 742

\bibitem[{{Iben} \& {Livio}(1993)}]{iben+livio93-1}
{Iben}, I.~J., {Livio}, M., 1993, PASP, 105, 1373

\bibitem[{{Koester}(2010)}]{koester10-1}
{Koester}, D., 2010, \memsai, 81, 921

\bibitem[{{Landsman} et~al.(1993){Landsman}, {Simon}, \&
  {Bergeron}}]{landsmanetal93-1}
{Landsman}, W., {Simon}, T., {Bergeron}, P., 1993, PASP, 105, 841

\bibitem[{{Livio} et~al.(1986){Livio}, {Truran}, \& {Webbink}}]{livioetal86-1}
{Livio}, M., {Truran}, J.~W., {Webbink}, R.~F., 1986, \apj, 308, 736

\bibitem[{{Loveridge} et~al.(2011){Loveridge}, {van der Sluys}, \&
  {Kalogera}}]{loveridgeetal11-1}
{Loveridge}, A.~J., {van der Sluys}, M.~V., {Kalogera}, V., 2011, \apj, 743, 49

\bibitem[{{Nebot G{\'o}mez-Mor{\'a}n} et~al.(2011)}]{nebotetal11-1}
{Nebot G{\'o}mez-Mor{\'a}n}, A., et~al., 2011, \aap, 536, A43

\bibitem[{{Paczynski}(1976)}]{paczynski76-1}
{Paczynski}, B., 1976, in {P.~Eggleton, S.~Mitton, \& J.~Whelan}, ed.,
  Structure and Evolution of Close Binary Systems, vol.~73 of \emph{IAU
  Symposium}, p.~75

\bibitem[{{Passy} et~al.(2012)}]{passyetal12-1}
{Passy}, J.-C., et~al., 2012, \apj, 744, 52

\bibitem[{{Rebassa-Mansergas} et~al.(2007){Rebassa-Mansergas}, {G{\"a}nsicke},
  {Rodr{\'{\i}}guez-Gil}, {Schreiber}, \&
  {Koester}}]{rebassa-mansergasetal07-1}
{Rebassa-Mansergas}, A., {G{\"a}nsicke}, B.~T., {Rodr{\'{\i}}guez-Gil}, P.,
  {Schreiber}, M.~R., {Koester}, D., 2007, MNRAS, 382, 1377

\bibitem[{{Rebassa-Mansergas} et~al.(2010){Rebassa-Mansergas}, {G{\"a}nsicke},
  {Schreiber}, {Koester}, \&
  {Rodr{\'{\i}}guez-Gil}}]{rebassa-mansergasetal10-1}
{Rebassa-Mansergas}, A., {G{\"a}nsicke}, B.~T., {Schreiber}, M.~R., {Koester},
  D., {Rodr{\'{\i}}guez-Gil}, P., 2010, \mnras, 402, 620

\bibitem[{{Rebassa-Mansergas} et~al.(2012){Rebassa-Mansergas}, {Nebot
  G{\'o}mez-Mor{\'a}n}, {Schreiber}, {G{\"a}nsicke}, {Schwope}, {Gallardo}, \&
  {Koester}}]{rebassa-mansergasetal12-1}
{Rebassa-Mansergas}, A., {Nebot G{\'o}mez-Mor{\'a}n}, A., {Schreiber}, M.~R.,
  {G{\"a}nsicke}, B.~T., {Schwope}, A., {Gallardo}, J., {Koester}, D., 2012,
  \mnras, 419, 806

\bibitem[{{Rebassa-Mansergas} et~al.(2011){Rebassa-Mansergas}, {Nebot
  G{\'o}mez-Mor{\'a}n}, {Schreiber}, {Girven}, \&
  {G{\"a}nsicke}}]{rebassa-mansergasetal11-1}
{Rebassa-Mansergas}, A., {Nebot G{\'o}mez-Mor{\'a}n}, A., {Schreiber}, M.~R.,
  {Girven}, J., {G{\"a}nsicke}, B.~T., 2011, \mnras, 413, 1121

\bibitem[{{Rebassa-Mansergas} et~al.(2008)}]{rebassa-mansergasetal08-1}
{Rebassa-Mansergas}, A., et~al., 2008, \mnras, 390, 1635

\bibitem[{{Ricker} \& {Taam}(2012)}]{ricker+taam12-1}
{Ricker}, P.~M., {Taam}, R.~E., 2012, \apj, 746, 74

\bibitem[{{Scargle}(1982)}]{scargle82-1}
{Scargle}, J.~D., 1982, ApJ, 263, 835

\bibitem[{{Schreiber} \& {G{\" a}nsicke}(2003)}]{schreiber+gaensicke03-1}
{Schreiber}, M.~R., {G{\" a}nsicke}, B.~T., 2003, A\&A, 406, 305

\bibitem[{{Schreiber} et~al.(2008){Schreiber}, {G{\"a}nsicke}, {Southworth},
  {Schwope}, \& {Koester}}]{schreiberetal08-1}
{Schreiber}, M.~R., {G{\"a}nsicke}, B.~T., {Southworth}, J., {Schwope}, A.~D.,
  {Koester}, D., 2008, \aap, 484, 441

\bibitem[{{Schreiber} et~al.(2010)}]{schreiberetal10-1}
{Schreiber}, M.~R., et~al., 2010, \aap, 513, L7+

\bibitem[{{Schwarzenberg-Czerny}(1996)}]{schwarzenberg-czerny96-1}
{Schwarzenberg-Czerny}, A., 1996, ApJ Lett., 460, L107

\bibitem[{{Soker} \& {Harpaz}(2003)}]{soker+harpaz03-1}
{Soker}, N., {Harpaz}, A., 2003, \mnras, 343, 456

\bibitem[{{Vennes} et~al.(1998){Vennes}, {Christian}, \&
  {Thorstensen}}]{vennesetal98-1}
{Vennes}, S., {Christian}, D.~J., {Thorstensen}, J.~R., 1998, \apj, 502, 763

\bibitem[{{Webbink}(1976)}]{webbink76-1}
{Webbink}, R.~F., 1976, \nat, 262, 271

\bibitem[{{Webbink}(1984)}]{webbink84-1}
{Webbink}, R.~F., 1984, ApJ, 277, 355

\bibitem[{{Webbink}(2008)}]{webbink07-1}
{Webbink}, R.~F., 2008, in {E.~F.~Milone, D.~A.~Leahy, \& D.~W.~Hobill}, ed.,
  Astrophysics and Space Science Library, vol. 352 of \emph{Astrophysics and
  Space Science Library}, p. 233

\bibitem[{{Willems} \& {Kolb}(2004)}]{willems+kolb04-1}
{Willems}, B., {Kolb}, U., 2004, A\&A, 419, 1057

\bibitem[{{Wood}(1995)}]{wood95-1}
{Wood}, M.~A., 1995, in {D.~Koester \& K.~Werner}, ed., White Dwarfs, vol. 443
  of \emph{Lecture Notes in Physics, Berlin Springer Verlag}, p.~41

\bibitem[{{Xu} \& {Li}(2010)}]{xu+li10-1}
{Xu}, X.-J., {Li}, X.-D., 2010, \apj, 716, 114

\bibitem[{{Zorotovic} et~al.(2010){Zorotovic}, {Schreiber}, {G{\"a}nsicke}, \&
  {Nebot G{\'o}mez-Mor{\'a}n}}]{zorotovicetal10-1}
{Zorotovic}, M., {Schreiber}, M.~R., {G{\"a}nsicke}, B.~T., {Nebot
  G{\'o}mez-Mor{\'a}n}, A., 2010, \aap, 520, A86

\bibitem[{{Zorotovic} et~al.(2011){Zorotovic}, {Schreiber}, \&
  {G{\"a}nsicke}}]{zorotovicetal11-1}
{Zorotovic}, M., {Schreiber}, M.~R., {G{\"a}nsicke}, B.~T., 2011, \aap, 536,
  A42

\end{thebibliography}

\end{document}